\documentclass[showpacs,amsmath,amssymb,aps,showkeys,floatfix,prd,a4paper]{revtex4}
\usepackage{subfigure}
\usepackage{dcolumn}
\usepackage{bm}
\usepackage{epsfig}
\usepackage{amsfonts}
\usepackage{amssymb,amscd}
\def\lsim{\raise0.3ex\hbox{$<$\kern-0.75em\raise-1.1ex\hbox{$\sim$}}}
\def\gsim{\raise0.3ex\hbox{$>$\kern-0.75em\raise-1.1ex\hbox{$\sim$}}}

\newcommand{\be}{\begin{equation}}
\newcommand{\ee}{\end{equation}}

\def\beq{\begin{equation}}
\def\eeq{\end{equation}}
\def\beqa{\begin{eqnarray}}
\def\eeqa{\end{eqnarray}}

\newcommand{\ba}{\begin{eqnarray}}   
\newcommand{\eea}{\end{eqnarray}}

\def\gappeq{\mathrel{\rlap {\raise.5ex\hbox{$>$}}
{\lower.5ex\hbox{$\sim$}}}}
\def\lappeq{\mathrel{\rlap{\raise.5ex\hbox{$<$}}
{\lower.5ex\hbox{$\sim$}}}}
\def\Toprel#1\over#2{\mathrel{\mathop{#2}\limits^{#1}}}

\begin{document}
\begin{flushright}
\vskip1cm
\end{flushright}

\title{Fully - heavy tetraquark production by $\gamma \gamma$ interactions \\ in hadronic collisions at the LHC}
\author{Victor P. Gon\c{c}alves}
\affiliation{High and Medium Energy Group, Instituto de F\'{\i}sica e Matem\'atica,  Universidade Federal de Pelotas,\\
Caixa Postal 354,  96010-900, Pelotas, RS, Brazil.}
\author{Bruno D.  Moreira}
\affiliation{Departamento de F\'isica, Universidade do Estado de Santa Catarina, 89219-710, Joinville, SC, Brazil. 
}
\begin{abstract}
We investigate the production of the fully - heavy tetraquark states $T_{4Q}$  in the  $\gamma \gamma$ interactions present  in proton-proton, proton-nucleus and nucleus-nucleus collisions at the CERN Large Hadron Collider (LHC). We focus on the $\gamma \gamma \rightarrow {\cal{Q}}{\cal{Q}}$  (${\cal{Q}} = J/\psi,\, \Upsilon$) subprocess, mediated by the $T_{4Q}$ resonance in the $s$ - channel,  and present predictions for the hadronic cross sections considering the kinematical ranges probed by the ALICE and LHCb Collaborations.  Our results demonstrate that the experimental study of this process is feasible and can be used  to investigate the existence and properties of the $T_{4c}(6900)$ and $T_{4b}(19000)$ states.
 \end{abstract}



\date{\today}

\maketitle


Over the last years the existence of exotic hadrons, which are a class of hadrons that cannot be easily accommodated in the remaining unfilled 
$q\bar{q}$ and $qqq$ states, has been established and a large number of candidates have been proposed (For recent reviews, see e.g. Refs. \cite{Karliner:2017qhf,Olsen:2017bmm,Liu:2019zoy}). In particular, the LHCb Collaboration has recently observed \cite{Aaij:2020fnh} a sharp peak in the di - $J/\psi$ channel at $M = 6.9$ GeV, which suggests the presence of a fully - charm tetraquark state. Such a result has motivated a series of studies about the description of the fully - heavy 
tetraquark states $T_{4Q}$, composed by charm and bottom quarks, which propose the existence of a large number of new exotic states (See e.g. Refs. 
\cite{Debastiani:2017msn,Bedolla:2019zwg,Giron:2020wpx,Chen:2020xwe,Chao:2020dml,Lu:2020qmp,liu:2020eha,Lu:2020cns,Wang:2020ols,Becchi:2020uvq,
Becchi:2020mjz,Karliner:2020dta,Weng:2020jao,Cao:2020gul,Yang:2020wkh,Wang:2020tpt}). In order to improve our understanding of these resonances, it is fundamental to have theoretical 
control of the mechanism in which they are produced. In recent years, different authors have proposed distinct production mechanisms of the $T_{4Q}$ in 
hadronic colliders (See e.g. Refs.   \cite{Karliner:2016zzc,Berezhnoy:2011xy,Carvalho:2015nqf,Esposito:2018cwh,Bai:2016int,Wang:2020gmd,Maciula:2020wri,Feng:2020riv,  Ma:2020kwb,Zhu:2020xni,Feng:2020qee}). In particular, in Ref. \cite{Carvalho:2015nqf}, the authors have 
proposed a model for the fully - charm tetraquark production in which the double $c\bar{c}$ pair is produced by the double scattering process 
and the cross section for $T_{4c}$ state is estimated within the framework of the color evaporation model \cite{cem}. Recently, such an idea was 
 elaborated in more detail in Ref. \cite{Maciula:2020wri}, which confirmed that this mechanism is one of the more promising ways to probe the $T_{4c}$ state.
  However, even this mechanism still has limitations due to the current theoretical uncertainties present in the description of the hadronization and double scattering processes. An important alternative to probe exotic hadrons, proposed and developed in recent years \cite{vicwer,vicmar,nosexotico,Klein:2019avl,Goncalves:2018hiw,Goncalves:2019vvo,Xie:2020ckr}, is the study of photon induced interactions at the LHC, which became a reality in the last decade (For a recent review see e.g. Ref. \cite{klein}). Our goal in this letter is to extend these previous studies for the $T_{4Q}$ production.  
  
\begin{figure}[t]
\centering
\includegraphics[scale=0.65]{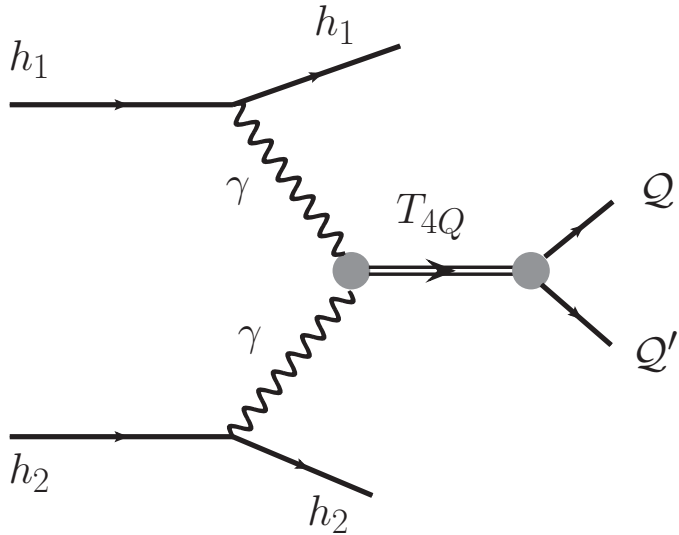}
\caption{Typical diagram for the production of a fully - heavy tetraquark  $T_{4Q}$ state by $\gamma \gamma$ interactions in a hadronic collision. The $T_{4Q} \rightarrow {\cal{Q}} {\cal{Q}}^{\prime}$ decay, where ${\cal{Q}}$ is a quarkonium state, is also represented.}
\label{esquema_colisao}
\end{figure}

One has that for ultra-relativistic collisions, the incident charged hadrons are an intense source of photons and in a collision at large impact parameters ($b > R_{h_1} + R_{h_2}$, with $R_i$ being the hadron radius), denoted hereafter ultra - peripheral collisions (UPCs),  photon -- photon and photon -- hadron  interactions become dominant over the strong hadron -- hadron one \cite{upc}. 
 In the particular case of the $T_{4Q}$ production by  photon - photon interactions,  represented in Fig. \ref{esquema_colisao}, the total cross section  can be factorized in terms of the equivalent flux of photons of the incident hadrons  and  the  photon-photon production cross section, which can be expressed in terms the two-photon decay width $\Gamma_{T_{4Q} \rightarrow \gamma \gamma}$. As the photon flux is well -- known, one has that this process is sensitive to the description of annihilation process, $T_{4Q} \rightarrow \gamma \gamma$, i.e. to the description of the $T_{4Q}$ wave function. Therefore, the study of the production in photon -- induced interactions allows us to directly test the modeling of the fully - heavy tetraquark states. Another advantage of the $T_{4Q}$ production by $\gamma \gamma$ interactions in hadronic colliders, is that the experimental separation of the associated events is relatively easy. As photon emission is coherent over the  hadron and the photon is colorless, the events will  be characterized by two intact recoiled hadrons (tagged hadrons) and the presence of two  rapidity gaps,  i.e., empty regions  in pseudo-rapidity that separate the intact very forward hadrons from the $T_{4Q}$ state,  which we will assume to decay into a pair of vector mesons. Such characteristics can be used to separate the events in a clean environment with a small background. In this letter we will perform an exploratory study, which will analyze the possibility of producing the exotic $T_{4c}$ and $T_{4b}$ states by two-photon interactions in $pp$, $pPb$ and $PbPb$ collisions at the LHC.  As we will show the resulting cross sections are large, which implies that a future experimental analysis is, in principle, feasible. Therefore, the study of this process can be used to confirm (or not) the existence  of these states as well it will allow us to investigate its properties.

Initially, let's present a brief review of the formalism needed to describe the $T_{4Q}$ tetraquark production by $\gamma \gamma$ interactions in hadronic collisions. Using the equivalent photon approximation \cite{epa,upc}, one has that the cross section for the collision between two hadrons, $h_{1}$ and $h_{2}$, is given by 
\begin{eqnarray}
\sigma \left( h_1 h_2 \rightarrow h_1 \otimes T_{4Q} \otimes h_2 ;s \right)   
&=& \int \hat{\sigma}\left(\gamma \gamma \rightarrow T_{4Q} ; 
W \right )  N\left(\omega_{1},{\mathbf b_{1}}  \right )
 N\left(\omega_{2},{\mathbf b_{2}}  \right ) S^2_{abs}({\mathbf b})  
\frac{W}{2} \mbox{d}^{2} {\mathbf b_{1}}
\mbox{d}^{2} {\mathbf b_{2}} 
\mbox{d}W 
\mbox{d}Y \,\,\, ,
\label{cross-sec-2}
\end{eqnarray}
where $\sqrt{s}$ is center-of-mass energy for the $h_1 h_2$ collision ($h_i$ = p, Pb), $\otimes$ characterizes a rapidity gap in the final state,  $W = \sqrt{4 \omega_1 \omega_2}$ is the invariant mass of the $\gamma \gamma$ system and $\omega_i$ are the photon energies. Moreover, $Y$ is the rapidity of the outgoing resonance $T_{4Q}$,  
 $N(\omega_i,b_i)$ is the equivalent photon spectrum generated by hadron (nucleus) $i$ at a distance $b_{i}$ from $h_i$ and the factor $S^2_{abs}({\mathbf b})$ is the absorption factor, which excludes the overlap between the colliding hadrons and allows to take into account only ultraperipheral collisions, where the impact parameter ${\mathbf b}$ is larger than the sum of the hadron radius. 
Finally, $ \hat{\sigma}_{\gamma \gamma \rightarrow T_{4Q}}(\omega_{1},\omega_{2})$ 
is the cross section for the production of a state $T_{4Q}$ from two real photons with energies $\omega_1$ and $\omega_2$. Using the Low formula \cite{Low}, the cross section for the production of  the $T_{4Q}$ 
state due to the two-photon fusion can be written in terms of the two-photon decay width $\Gamma_{T_{4Q} \rightarrow \gamma \gamma}$ as  follows
\begin{eqnarray}
 \hat{\sigma}_{\gamma \gamma \rightarrow T_{4Q}}(\omega_{1},\omega_{2}) = 
8\pi^{2} (2J+1) \frac{\Gamma_{T_{4Q} \rightarrow \gamma \gamma}}{M_{T_{4Q}}} 
\delta(4\omega_{1}\omega_{2} - M_{T_{4Q}}^{2}) \, ,
\label{Low_cs}
\end{eqnarray}
where $M_{T_{4Q}}$ and $J$ are, respectively, the mass and spin of the  produced  state. As in Ref. \cite{Goncalves:2018hiw}, we will estimate the photon flux assuming that  nucleus (proton) can be described by  a monopole (dipole) form factor and that 
$S^2_{abs}({\mathbf b}) = \Theta\left(
\left|{\mathbf b}\right| - R_{h_1} - R_{h_2}
 \right )  = 
\Theta\left(
\left|{\mathbf b_{1}} - {\mathbf b_{2}}  \right| - R_{h_1} - R_{h_2}
 \right ) $, 
where $R_{h_i}$ is the radius of the hadron $h_i$ ($i = 1,2$), with $R_p = 0.7$ fm and $R_{A} = 1.2 \, A^{1/3}$ fm.  
A detailed discussion about the  theoretical uncertainty associated to these choices is presented in Ref. \cite{celsina}.

One has that the cross section is directly dependent on the values for  
the decay width $\Gamma_{T_{4Q} \rightarrow \gamma \gamma}$, mass and spin of the resonance. Such quantities can  be taken from experiment or can be theoretically estimated. In our analysis we will consider  that the resonance decays into a 
${\cal{Q}} {\cal{Q}}^{\prime}$ final state, where ${\cal{Q}}$ is a quarkonium state. As a consequence, the cross section will be proportional to $\Gamma_{T_{4Q} \rightarrow \gamma \gamma} \times {\cal{B}}(T_{4Q} \rightarrow {\cal{Q}} {\cal{Q}}^{\prime})$, where 
${\cal{B}}(T_{4Q} \rightarrow {\cal{Q}} {\cal{Q}}^{\prime})$ is the associated branching fraction. 
In principle, such a product can be measured precisely in future $ e^+ e^-$ colliders, which will make  the predictions for the LHC will be parameter free. However,  as these quantities are currently unknown, we will assume some naive approximations in order to derive an estimate of the associated cross sections. We will focus on the case in that ${\cal{Q}} =  {\cal{Q}}^{\prime}$ and ${\cal{Q}}  = J/\psi$ or $\Upsilon$, which are expected to be present in the final state when the $T_{4c}$ and $T_{4b}$ are produced, respectively. 
Moreover, we will assume that the fully - charm tetraquark $T_{4c}$ is the $X(6900)$ state, recently observed by the LHCb Collaboration \cite{Aaij:2020fnh}.  On the other hand, for the $T_{4b}$ case, we will assume that it corresponds to the $X(19000)$ tetraquark state, predicted by different phenomenological models (See e.g. Refs. \cite{Lu:2020qmp,Lu:2020cns,Becchi:2020mjz,Karliner:2020dta,Esposito:2018cwh,Bai:2016int}). As the quantum numbers of these resonances  are still unknown,  we will estimate the cross sections assuming that  $J^P = 0^+$ or $2^+$. Motivated by the LHCb results \cite{Aaij:2020fnh} and following Ref. \cite{Wang:2020gmd}, we will take ${\cal{B}}(T_{4Q} \rightarrow {\cal{Q}} {\cal{Q}}) = 2\%$ (For a recent calculation of the $T_{4c}$ branching ratio see, e.g., Ref. \cite{Chen:2020xwe}). Moreover, as in our case the $T_{4Q}$ state is expected to have spin - parity identical to that from the $\chi_Q$ quarkonium family, we will assume that 
$\Gamma_{T_{4Q} \rightarrow \gamma \gamma} \simeq \Gamma_{\chi_Q \rightarrow \gamma \gamma}$. Therefore, our predictions for the $X(6900)$ production will be derived using the values for  $\Gamma_{\chi_c \rightarrow \gamma \gamma}$ presented in the latest Particle Data Group \cite{pdg}.  On the other hand, as the two - photon width $\Gamma_{\chi_b \rightarrow \gamma \gamma}$ was not still measured, we will estimate the $X(19000)$ production assuming the values derived in Ref. \cite{Hwang:2010iq} using the covariant light - front framework. It is important to emphasize that our results can be easily 
generalized to other values of the branching ratio and two - photon width by 
a simple rescaling of our predictions, since they are linearly dependent on 
these quantities.

\begin{table}[t] 
\centering
\begin{tabular}{||c|c|c|c|c||} 
\hline 
\hline
Collision & Resonance & LHC &  LHCb  &  ALICE \\
\, & \, & Full rapidity range & $2.0 \le Y \le 4.5$ &  $-1.0 \le Y \le 1.0$ \\
\hline
\hline
$pp$ ($\sqrt{s} = $13 TeV) &  $X(6900)$, $0^{++}$   & 26.3 fb & 	5.53	fb  & 6.34 fb     \\  
 \,                        &   $X(6900)$, $2^{++}$  & 31.9 fb &	6.71	fb  & 7.71 fb   \\ 
\hline
\hline
$p Pb$  ($\sqrt{s} = $8.1 TeV) &   $X(6900)$, $0^{++}$ & 76.3 pb &  21.6 pb  & 22.4 pb \\  
  \,                         &   $X(6900)$, $2^{++}$ & 92.4 pb & 26.2 pb  & 27.2 pb   \\ 
\hline
\hline
$PbPb$ ($\sqrt{s} = $5.02 TeV) & $X(6900)$, $0^{++}$     & 171.0 nb & 22.3 nb      & 70.0 nb     \\  
  \,                           &    $X(6900)$, $2^{++}$  & 206.0 nb &  26.7 nb      & 84.7 nb    \\ 
\hline
\hline
\end{tabular}
\caption{Total cross sections for the $X(6900)[J^P] \rightarrow J/\psi J/\psi$ production by $\gamma \gamma$ interactions in $pp$, $pPb$ and $PbPb$ collisions for different center - of - mass energies considering the  full LHC rapidity range as well as the rapidity ranges covered by the  ALICE and LHCb detectors.} 
\label{tab:x6900}
\end{table}

\begin{figure}[t]
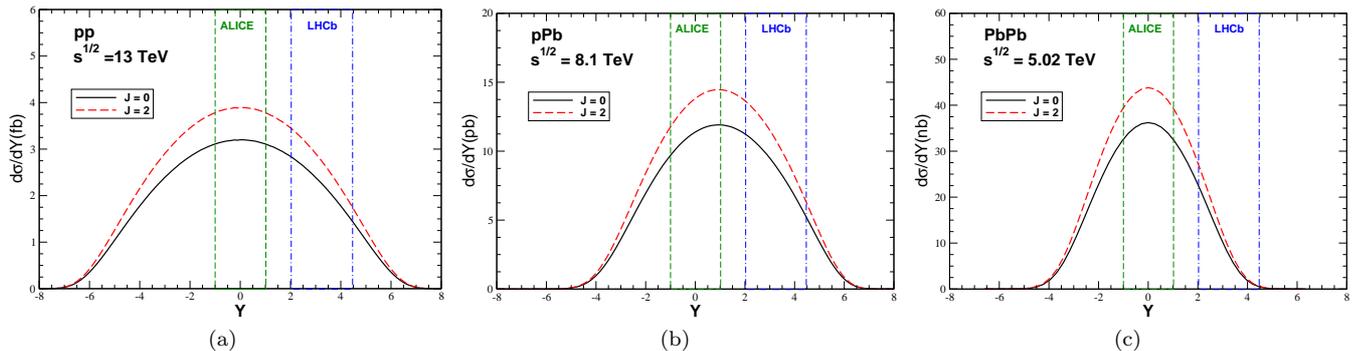

\begin{center}
\subfigure[ ]{\label{figa}
\includegraphics[width=0.32\textwidth]{X6900_pp.eps}}
\subfigure[ ]{\label{figb}
\includegraphics[width=0.32\textwidth]{X6900_pPb.eps}}
\subfigure[ ]{\label{figc}
\includegraphics[width=0.32\textwidth]{X6900_PbPb.eps}} 
\end{center}
\caption{
Rapidity distributions for the $X(6900) \rightarrow J/\psi J/\psi$ production by $\gamma \gamma$ interactions in (a) $pp \,(\sqrt{s} = 13 \,\mbox{TeV})$, (b) $pPb \, (\sqrt{s} = 8.1\,\mbox{TeV})$ 
and (c) $PbPb \, (\sqrt{s}=5.02\,\mbox{TeV})$ collisions at the LHC.
}
\label{fig:rap6900}
\end{figure}

In Table \ref{tab:x6900} we present our predictions for the total cross sections for the $X(6900) \rightarrow J/\psi J/\psi$ production in $pp/pPb/PbPb$ collisions at the LHC energies considering the full LHC rapidity range as well as the rapidity ranges covered by the ALICE and LHCb detectors. We consider the two possible values for $J^P$ and, following Ref. \cite{Karliner:2020dta},   we will assume  that the mass of the resonance is equal to { {6871.0 MeV for $J=0$ and 6967.0 MeV for $J=2$}}.  Due to the $Z^2$ dependence of the photon spectra, we have that the following hierarchy is approximately valid  for the $X(6900)$ production induced by $\gamma \gamma$ interactions: $\sigma_{PbPb} = Z^2 \cdot \sigma_{pPb} = Z^4 \cdot \sigma_{pp}$, with $Z = 82$.  In Fig. \ref{fig:rap6900} the corresponding rapidity distributions for the $X(6900) \rightarrow J/\psi J/\psi$ production  are presented.  One has that the predictions for the  $J = 2$ resonance are larger than those for the $ J = 0$ one, as expected from the Low formula.  Due to the asymmetry in the proton and nuclear photon fluxes present in the initial state, we predict an asymmetric rapidity distribution in the case of $pPb$ collisions. In Fig. \ref{fig:rap6900} we also indicate the kinematical rapidity ranges probed by the ALICE ($-1 \le Y \le +1$) and LHCb ($+2 \le Y \le +4.5$) detectors.  The resulting predictions for the total cross sections in the ALICE and LHCb rapidity ranges are also  presented in Table \ref{tab:x6900}. In comparison with the results for the full LHC rapidity range, one has that the predictions are reduced by a factor between $3.0$ and $8.0$ depending on the initial state and the rapidity range covered by the detector, with the larger reduction being for $PbPb$ collisions at the LHCb.  Although this reduction is nonnegligible, the final values  are still large and imply a significant number of events if we consider that the expected integrated luminosity for the high luminosity run of the LHC is 50/fb (10/nb)  for $pp$ ($PbPb$) collisions \cite{hl_fcc}. 
In particular, we predict  that the number of events per year in $pp$ ($PbPb$) collisions will be $\approx$  276 (335) at LHCb and 385 (840) at ALICE, which implies that the experimental analysis of this process is, in principle, feasible.

The analysis performed above can be directly extended for the production of fully - bottom  tetraquark states. In particular, we will  provide predictions for the $X(19000)$ production, assuming that { {M = 19434.0 MeV for J=0 and $M = 19481.0$ MeV for $J=2$ as predicted in Ref. \cite{Karliner:2020dta}}}. Our results for the cross sections are presented in Table \ref{tab:x19000}. In comparison to the results for the $X(6900)$, the cross sections for the 
$X(19000)[J^P] \rightarrow \Upsilon \Upsilon$ production are smaller by a factor $\gtrsim 10^3$. As a consequence, the associated number of events per year will be very small, making the  experimental analysis of this exotic state in the next run of the LHC a hard task. However, one has verified that in $PbPb$ collisions at $\sqrt{s} = 39$ TeV, which is the expected center - of - mass energy  of the Future Circular Collider (FCC) \cite{fcc}, the cross sections are increased by a factor $\approx 3$. As the integrated luminosity for this future collider is expected to be $\approx$ 110/nb, one has that the analysis of the  $X(19000)[J^P] \rightarrow \Upsilon \Upsilon$ production in ultraperipheral $PbPb$ collisions  becomes feasible at the FCC.

\begin{table}[t] 
\centering
\begin{tabular}{||c|c|c|c|c||} 
\hline 
\hline
Collision & Resonance & LHC &   LHCb  &  ALICE \\
\, & \, & Full rapidity range & $2.0 \le Y \le 4.5$ &  $-1.0 \le Y \le 1.0$ \\
\hline
\hline
$pp$ ($\sqrt{s} = $13 TeV) &  $X(19000)$, $0^{++}$      & 2.40$\times10^{-3}$ fb & 4.90$\times10^{-4}$ fb &   6.88$\times10^{-4}$ fb \\  
 \,                        &   $X(19000)$, $2^{++}$     & 5.91$\times10^{-3}$ fb & 1.21$\times10^{-3}$ fb &   1.70$\times10^{-3}$ fb  \\ 
\hline
\hline
$pPb$  ($\sqrt{s} = $8.1 TeV) &   $X(19000)$, $0^{++}$    & 5.60 fb & 1.62 fb & 1.96 fb  \\  
  \,                         &   $X(19000)$, $2^{++}$    & 13.80 fb & 3.99 fb  &  4.83 fb   \\ 
\hline
\hline
$PbPb$ ($\sqrt{s} = $5.02 TeV) & $X(19000)$, $0^{++}$     & 8.33 pb & 0.564 pb &   4.32 pb   \\  
  \,                           &    $X(19000)$, $2^{++}$  & 20.5 pb & 1.38 pb &   10.6 pb  \\ 
\hline
\hline
\end{tabular}
\caption{Total cross sections for the $X(19000)[J^P] \rightarrow \Upsilon \Upsilon$ production by $\gamma \gamma$ interactions in $pp$, $pPb$ and $PbPb$ collisions for different center - of - mass energies considering the  full LHC rapidity range as well as the rapidity ranges covered by the  ALICE and LHCb detectors.} 
\label{tab:x19000}
\end{table}


Some comments are in order. As discussed before, the $T_{4Q}$ production by $\gamma \gamma$ interactions in 
$pp/pPb/PbPb$ collisions will be characterized by two intact hadrons that can be detected by forward detectors and two rapidity gaps in the final state. This final state is also generated by the single and double scattering mechanisms discussed in Refs. \cite{Goncalves:2002vq,wolfgang,bruno_duplo,bruno_dps}. The comparison of our results for the 
$X(6900) \rightarrow J/\psi J/\psi$ case with those derived in Refs. \cite{bruno_duplo,bruno_dps} for the double $J/\psi$ production   indicates that both processes have similar cross sections before the inclusion of additional kinematical cuts. Such a result strongly motivates  a more detailed analysis, including the cuts usually considered by the experimental collaborations. Another important comment is that during the development of this analysis we were aware that a similar study of the $X(6900)$ production in UPHICs  is being independently performed by Y. Xie and collaborators \cite{yapingx}. These authors have estimated the decay width of $X(6900)$ to two photons using the effective Lagrangian method and derived slightly smaller values for the cross sections. Such result indicates that the naive assumptions assumed in our analysis are a good first approximation to estimate the fully - heavy tetraquark production by $\gamma \gamma$ interactions in hadronic collisions.

Finally, let's summarize our main results and conclusions. Motivated by the observation of  a sharp peak in the di - $J/\psi$ channel at $M = 6.9$ GeV by 
the LHCb Collaboration,  which suggest the presence of a fully - charm tetraquark state, we have developed in this letter the treatment of the fully - heavy tetraquark production by $\gamma \gamma$ interactions in hadronic collisions. We have focused on the $X(6900)$ and $X(19000)$ states, and estimated the cross sections assuming that they decay into a $J/\psi J/\psi$ and $\Upsilon \Upsilon$ final state, respectively. Predictions for the kinematical ranges probed by the ALICE and LHCb detectors were presented, which  indicate that the experimental analysis of the $X(6900)$ production  is, in principle, feasible at the LHC. Such conclusion strongly motivates a more detailed analysis, taking into account of the experimental cuts assumed by the distinct experimental collaborations, which we intend to perform in a near future.

\begin{acknowledgments}
One of the authors (VPG) acknowledges useful discussions about 
the subject with Ya-Ping Xie.  This work was  partially financed by the Brazilian funding agencies CNPq,  FAPERGS and  INCT-FNA (process number 464898/2014-5).
 \end{acknowledgments}

\hspace{1.0cm}

\end{document}